\documentclass[%
 reprint,prc,
amsmath,amssymb,
aps,
floatfix,
]{revtex4-1}

\usepackage{graphicx}
\usepackage{dcolumn}
\usepackage{bm}
\usepackage{url}
\usepackage[colorlinks=true,linktocpage=true,linkcolor=blue,citecolor=blue,allcolors=
blue] {hyperref }
\usepackage[mathlines]{lineno}
\usepackage{times} 
\usepackage{lipsum}

\begin{document}
\preprint{APS/123-QED}

\title{Probing vorticity and fluctuations in a rotating hadron resonance gas at LHC energy}
\author{S. Ipsita Sahoo$^{1,2}$}%
\email{sisahoo@barc.gov.in}
\author{S. Biswal$^{2}$}
\email{sanjeebani17@gmail.com}
\author{D. Dutta$^{1,2}$}
\email{ddutta08@hbni.ac.in}
\author{D. K. Mishra$^{1,2}$}
\email{dkmishra@barc.gov.in}

\affiliation{$^1$Bhabha Atomic Research Centre, Mumbai - 400085, INDIA}
\affiliation{$^2$Homi Bhabha National Institute, Anushaktinagar, Mumbai - 400094, INDIA}

\date{\today}

\begin{abstract}
  A large vorticity produced in non-central ultra-relativistic
  heavy-ion collisions induces an effective chemical potential in both
  partonic and hadronic matter, thereby influencing the quark-hadron
  transition and its associated properties. In this work, we investigate the influence of rotation
  on hadron yields within the framework of Hadron Resonance Gas (HRG) model.
  Our results show that vorticity significantly modifies hadron yields and
  their ratios. Most  notably, rotation enhances the $p/\pi^+$ ratio while
  suppressing the $K^+/\pi^+$ ratio, suggesting that these observables may serve
  as sensitive probes of the rotational properties of the medium
  created in heavy-ion collisions.
  To quantify the effect of rotation, the calculated dependence of the
  $p/\pi^+$ ratio on vorticity is compared with the centrality dependence
  of the $p/\pi$ ratio measured by the ALICE collaboration in Pb+Pb
  collisions at $\sqrt{s_{_{NN}}}$ = 5.02 TeV.
  From this comparison, we estimate the maximum vorticity produced at freeze-out
  for different collision centralities. In case of peripheral collisions,
  the freeze-out vorticity ($\omega$) is found to reach an upper bound value of
  approximately 0.088 GeV, whereas for central collisions it is
  about 0.028 GeV. Furthermore, we investigate the effect of rotation on
  fluctuations of conserved quantities and their correlations. This study provides a 
  quantitative framework for assessing the role of rotation in the thermodynamics of 
  hadronic matter and its phenomenological consequences for heavy-ion collisions.
  
\end{abstract}

\maketitle

\section{Introduction}\label{intro} 
The properties of the fireball generated in relativistic heavy-ion collisions 
have been extensively investigated over the past several decades through both 
experimental measurements and theoretical studies. A central objective of the 
research programs at the Relativistic Heavy Ion Collider (RHIC) and the Large 
Hadron Collider (LHC) is to probe the behavior of Quantum Chromo-dynamics (QCD) 
matter under extreme temperature and energy-density
conditions~\cite{Bzdak:2019pkr,Rischke:2003mt}. 
Under such conditions, the emergent state of matter-the Quark-Gluon Plasma 
(QGP), which exhibits collective behavior resembling that of a nearly perfect fluid, 
is characterized by an exceptionally small shear viscosity-to-entropy density 
ratio ($\eta/s$). 

In non-central relativistic heavy-ion collisions, the resulting fireball can 
sustain rapid rotation, generating a large angular momentum that induces vorticity in the
fluid~\cite{Otsuka:2019diq,Fukushima:2019,Huang2021,Deng:2016}. In such 
events, the total angular momentum ($J$) carried by the colliding nuclei is 
proportional to both the impact parameter, $b$, and the center-of-mass energy 
($\sqrt{s_{NN}}$) following the relation, $J\propto b\sqrt{s_{NN}}$. 
Although a significant fraction of this angular momentum is carried away by the 
spectator nucleons, a finite portion remains within the fireball, thereby 
inducing finite fluid vorticity within the evolving medium~\cite{Deng:2016}.
Consequently, the 
fireball develops a local angular velocity of the order of $0.01$–$0.1$ 
GeV~\cite{LiangWang:2016,JiangLiao:2016prl}. The local vorticity
gradually decreases 
during the hydrodynamic expansion of the system, but if the vorticity remains sufficiently 
large until the kinetic freeze-out stage, it can be measured experimentally as an observable.

In this context, the STAR Experiment has provided the first direct experimental 
observation of finite 
global hyperon polarization in relativistic heavy-ion collisions, providing 
compelling evidence for the existence of finite vorticity in the medium, 
estimated to be of the order of $10^{21}~\mathrm{s}^{-1}$, representing the most 
vortical fluid ever observed in any physical system~\cite{STAR:2017ckg}. 
Such extreme vorticity can significantly influence the dynamical evolution of 
the system and modify the thermodynamic and transport properties of strongly 
interacting QCD matter~\cite{Fujimoto:2021xix}. These transport properties, 
particularly shear viscosity, play a crucial role in the generation and evolution of finite 
vorticity, indicating a strong interplay between dissipative effects and 
rotational dynamics in the system~\cite{Huang:2015oca,Otsuka:2019diq,Bali:2011qj}.
These two phenomena are intrinsically 
coupled, as dissipative effects significantly modulate the generation and 
persistence of fluid angular momentum~\cite{Fu:2021pok}.

In addition to large vorticity, the fireball produced in relativistic heavy-ion 
collisions is subjected to extremely intense electromagnetic fields, 
particularly strong magnetic fields generated during the early stages of 
non-central collisions~\cite{Fukushima:2008xe}. Recent studies demonstrate
that the magnetic field 
produced in such collisions can significantly influence hadron yields and 
conserved-charge fluctuations, thereby directly affecting the chemical freeze-out 
conditions of the medium~\cite{Mukherjee:2023ijv,Vovchenko:2024wbg,Padhan:2026mwg}.
Consequently, in addition to 
temperature ($T$) and baryon chemical potential ($\mu_B$), both the magnetic 
field ($B$) and angular velocity (vorticity, $\omega$) emerge as  
important thermodynamic control parameters governing the properties and phase 
structure of strongly interacting QCD matter~\cite{Mukherjee:2023qvq,Ebihara:2017,Fukushima:2019}.

In rotating QCD matter, several intriguing phenomena may emerge, including the 
chiral vortical effect and the chiral vortical wave, which are induced by the 
rotational motion of the fluid~\cite{Son:2009}. Recently, significant
attention has been devoted 
to the study of QCD in rotating reference frames through lattice simulations 
\cite{Yamamoto:2013zwa}. This progress further motivates the use of 
QCD-inspired models to investigate the properties and dynamics of the dense
rotating QCD matter~\cite{Chen:2015hfc,JiangLiao:2016prl,Ebihara:2017,
  Chernodub:2017,Wang:2018sur}.

The rotational motion of the medium couples non-trivially with both the 
magnetic field and the thermal background, leading to substantial modifications 
in the dynamical evolution of the system. Therefore, a consistent description of 
the medium created in ultra-relativistic heavy-ion collisions necessitates the 
inclusion of rotational effects. From a thermodynamic perspective, rotation 
effectively introduces an additional chemical-potential-like contribution to the 
system, which can alter the phase structure and influence the nature of the QCD 
phase transition. In this regard, investigating the role of rotation in 
understanding the QCD phase diagram and properties of strongly interacting
matter under extreme conditions is of considerable theoretical and 
phenomenological interest.

The thermodynamic properties of hot and dense QCD matter are widely investigated 
within the framework of the hadron resonance gas (HRG) model, which has proven to be
highly successful in describing hadron yields and their respective ratios, for 
the extraction of chemical freeze-out parameters, namely the temperature $T$ 
and baryon chemical potential $\mu_B$, in heavy-ion collisions 
\cite{Cleymans:2005xv}. The HRG model provides a reliable description of 
strongly interacting matter at temperatures below the QCD transition 
temperature; however, at higher temperatures, its predictions progressively 
deviate from lattice QCD calculations~\cite{Bazavov:2012vg}. In its
conventional formulation, the 
medium is approximated as a gas of non-interacting, point-like hadrons 
and resonances, thereby neglecting finite-width effects and residual 
interactions among the constituents.

Motivated by the significant role of rotation in the medium produced during 
non-central heavy-ion collisions, we systematically investigate the influence of 
rotation on hadron abundances and their yield ratios within the HRG framework. 
In particular, we study the centrality dependence of the proton-to-pion ratio, 
$p/\pi$, as a potential observable sensitive to rotational effects at chemical 
freeze-out in heavy-ion collisions. An attempt is made to estimate the upper limit of 
the rotation ($\omega$) using the experimental data. Further, we also present the effect of
rotation on fluctuation observables and their correlations~\cite{Vovchenko:2024wbg,Marczenko:2024kko}.

The manuscript is organized as follows: Section~\ref{intro} provides the 
introduction and motivation of the present study; Section~\ref{hrg} outlines the 
formulation of the HRG model to incorporate rotational effects; 
Section~\ref{results} discusses the effects of rotation on hadron yields, 
their ratios, susceptibilities and correlations; Section~\ref{summary} concludes with a summary of our 
findings.

\section{HRG model with rotation}\label{hrg}

The Hadron Resonance Gas (HRG) model describes the confined phase of QCD matter 
as a grand canonical ensemble of non-interacting relativistic gas composed of 
all experimentally established hadrons and resonant
states~\cite{Andronic:2012ut,Karsch:2011,Garg:2013,Braun:1995,Gupta:2022}.
Within the hadronic 
medium, the precise identification of the relevant degrees of freedom
and their effective interactions is essential for an accurate thermodynamic 
description. The HRG framework treats the system as a multi-component ideal gas, 
accounting for the full spectrum of known baryons and mesons. In the present 
work, we extend this formal treatment by incorporating rotational effects into 
the ideal HRG formalism.

The pressure corresponding to the $i^{\mathrm th}$ baryonic ($b$) or mesonic 
($m$) species is expressed as~\cite{Fujimoto:2021xix,Mukherjee:2023qvq,Chen:2015hfc}:

\begin{equation}\label{e1}
\begin{split}
     p_i^{b/m}=\pm\frac{T}{8\pi^2}\sum_{l=-\infty}^{\infty}\int dk_r^2 
     \int dk_z \sum_{\nu=l}^{l+2s_i} J_\nu^2(k_r r) \\ \times \ln(1\pm 
e^{-(E_{l,i}-\mu_i)/T}).
\end{split}
\end{equation}

Here, $\mu_i=B_i\mu_B+Q_i\mu_Q+S_i\mu_S$ denotes the total chemical potential of 
the $i^{\mathrm{th}}$ hadronic species, where $B_i$, $Q_i$, and $S_i$ represent 
its baryon number, electric-charge and  strangeness quantum number, 
respectively. The quantities $\mu_B$, $\mu_Q$, and $\mu_S$ representing their 
associated chemical potentials. The term $J_\nu^2(k_r r)$ represents the 
squared Bessel function of the first kind arising from the cylindrical symmetry 
of the rotating system~\cite{Chernodub:2020qah}.
In the presence of rotation, the single-particle energy spectrum, $E_{l,i}$, is 
expressed as: 
\begin{equation}\label{e2}
    E_{l,i}=\sqrt{k_r^2+k_z^2+m_i^2}-(l+s_i)\omega,
\end{equation}
where $m_i$ and $s_i$ denote the rest mass and intrinsic spin of the 
$i^{\mathrm th}$ particle, respectively and $l$ represents the orbital angular 
momentum quantum number. The second term acts as an effective chemical potential due to rotation.

The hadron number densities $n_i$, obtained from the derivative of pressure 
with respect to the corresponding chemical potential
$\mu_i$, is given by,

\begin{equation}\label{e2}
    \begin{split}
        n_i^{b/m}=\frac{1}{8\pi^2}\sum_{l=-\infty}^{\infty}\int dk_r^2 \int dk_z \sum_{\nu=l}^{l+2s_i} J_{\nu}^2 (k_r r) \\
        \times \frac{1}{e^{(E_{l,i}-\mu_i)/T}\pm 1}.
    \end{split}
\end{equation}

The causality constraint prohibits the formation of unphysical condensate 
states by requiring that the rotational contribution, $(l+s)\omega$, does not 
exceed the `free-particle' component of the energy dispersion 
$\sqrt{k_r^2+k_z^2+m_i^2}$. By imposing a normalization boundary condition 
within a finite cylindrical volume $r\le R$, the causality requirement reduces 
to $R\omega \le 1$. Furthermore, the causality bound induces a discretization of 
the momentum ($k_r$) modes in the low-momentum region, leading to the 
introduction of an infrared cutoff in the transverse momentum integration, 
defined as
$\Lambda_l^{IR}=\zeta_{l,1}\omega$. Here, $\zeta_{l,1}$ denotes the first zero 
of the Bessel function, defined by $J_l(\zeta_{l,1})=0$. Consequently, the 
$k_r$ integration in equations \ref{e1} and \ref{e2} is modified as, 
\begin{equation}\label{e3}
    \int dk_r^2 \rightarrow \int_{(\Lambda_l^{IR})^2} dk_r^2.
\end{equation}

\begin{figure}[htbp]
    \centering
    \includegraphics[width=\linewidth]{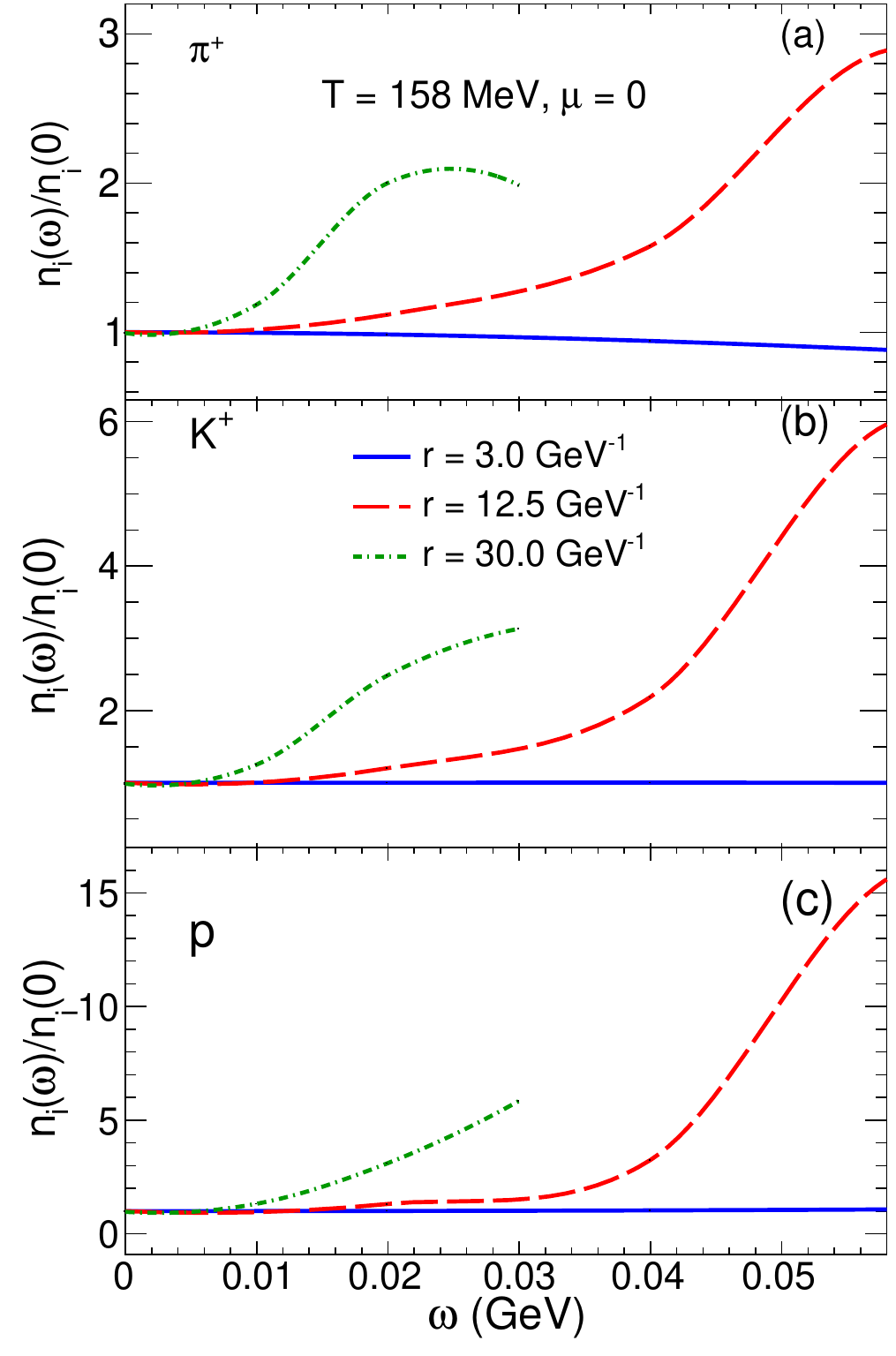}
    \caption{Variation of number densities for $\pi^+$ (top), $K^+$ (middle), 
and $p$ (bottom) with rotation for different $r$ values.}
    \label{fig1}
\end{figure}

In Heavy-Ion Collisions (HIC), the geometry of the ``fireball" depends 
significantly on the collision centrality. For central collisions, the fireball 
is approximated as nearly spherical with a radius of $R=30$ GeV$^{-1}$ or $6$ 
fm. For peripheral collisions, the maximum radius is taken as $R=12.5$ 
GeV$^{-1}$ or $2.5$ fm, since the overlap region between the two colliding 
nuclei takes on a spindle or almond-like shape~\cite{STAR:2017sal}.
Figure~\ref{fig1} shows the 
dependence of the normalized hadron number densities ($\pi^+$, $K^+$, and $p$) 
with rotation for different radial distances $r$ = $3$, $12.5$, and $30$ 
GeV$^{-1}$. 
In the present analysis, we fix the value of $r$ to be $3$ GeV$^{-1}$ to allow 
access to a wider range of vorticity $\omega$.

\section{Results and Discussions}\label{results}

The study investigates the effects of rotation ($\omega$) on hadron yields and 
their respective ratios. Previously, the enhancement of the $p/\pi$ ratio and 
suppression of the $n/\pi$ ratio were identified as possible signatures of 
magnetic field effects~\cite{Vovchenko:2024wbg}. Consequently, we explore whether 
those observables exhibit similar sensitivity in presence of rotation, which has effects 
analogous to that of a magnetic field.

\subsection{Effect of rotation on hadron yields and their ratios}
The rotation affects the effective chemical potential of 
hadrons through its coupling to the orbital angular momentum and spin of 
the particles, given by \(\mu_{\mathrm {eff}}=(l+s_z)\omega\). This coupling 
alters the hadron yields as well as the thermodynamic observables. In a rotating 
medium, this interaction induces a shift in the single-particle energy 
spectrum, fundamentally altering the phase-space distribution functions. In 
the present analysis, we include all hadrons of mass upto $2.5$ GeV listed by the Particle Data 
Group~\cite{PDG:2014}.

\begin{figure}[htbp]
    \centering
    \includegraphics[width=\linewidth]{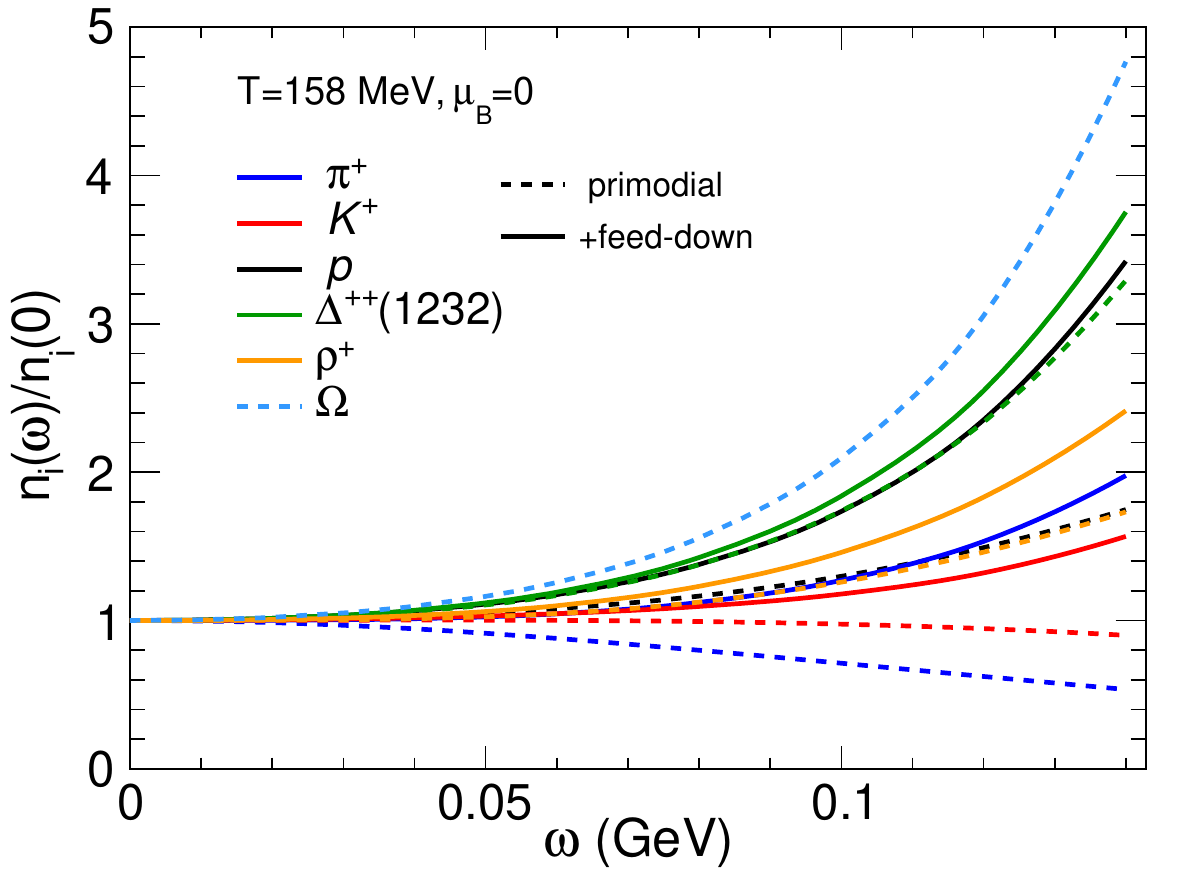}
    \caption{Dependence of various hadron yields (normalized by their 
corresponding values without rotation) on rotation, $\omega$, in HRG model at 
temperature $T=158$ MeV and vanishing chemical potential. The dashed lines 
correspond to the primordial hadron abundances, while the solid lines 
incorporate strong and electromagnetic decay feed down.}
    \label{fig2}
\end{figure}

\begin{figure}[htbp]
    \centering
    \includegraphics[width=\linewidth]{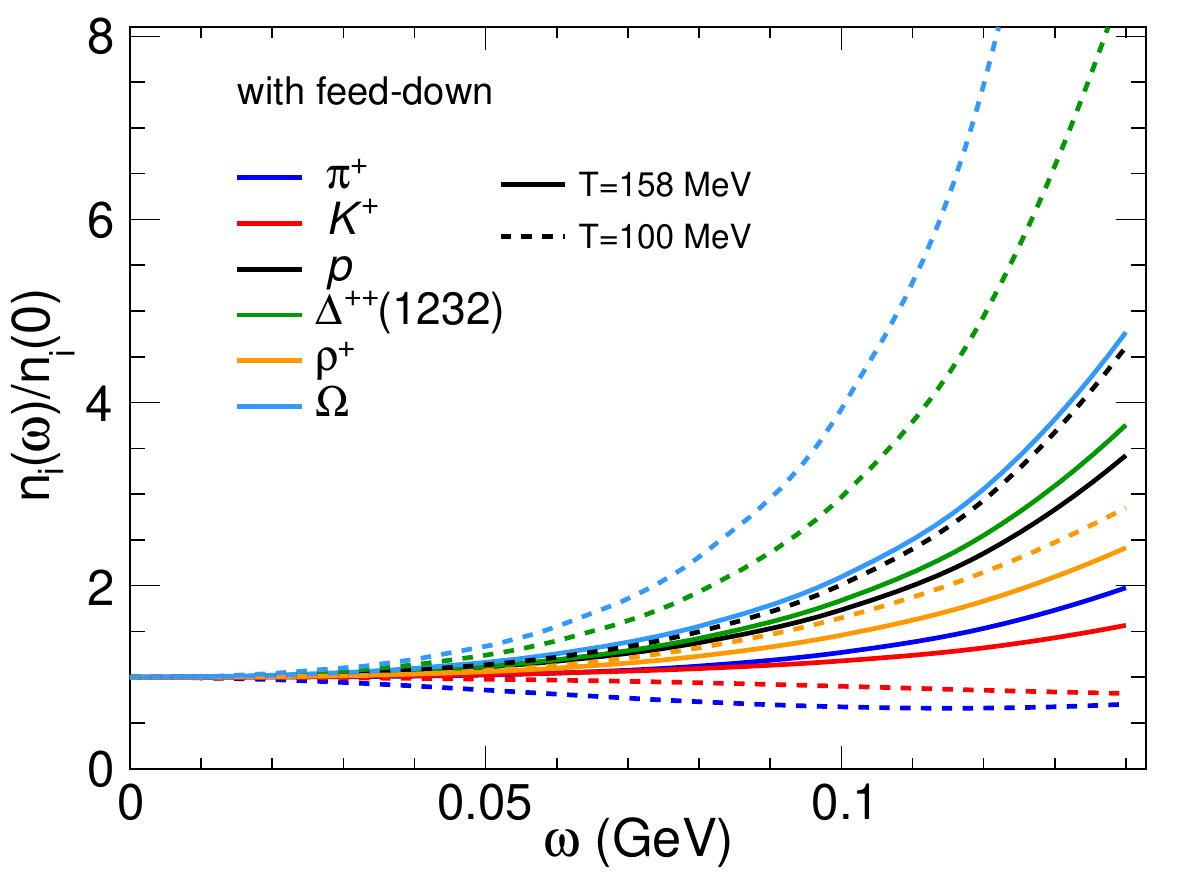}
    \caption{Particle densities normalized to their corresponding densities without 
    rotation as a function of rotation, $\omega$, for two different temperatures
      $T$ = $158$ MeV and $100$ MeV, at vanishing chemical potential.}
    \label{fig3}
\end{figure}

\begin{figure}[htbp]
    \centering
    \includegraphics[width=\linewidth]{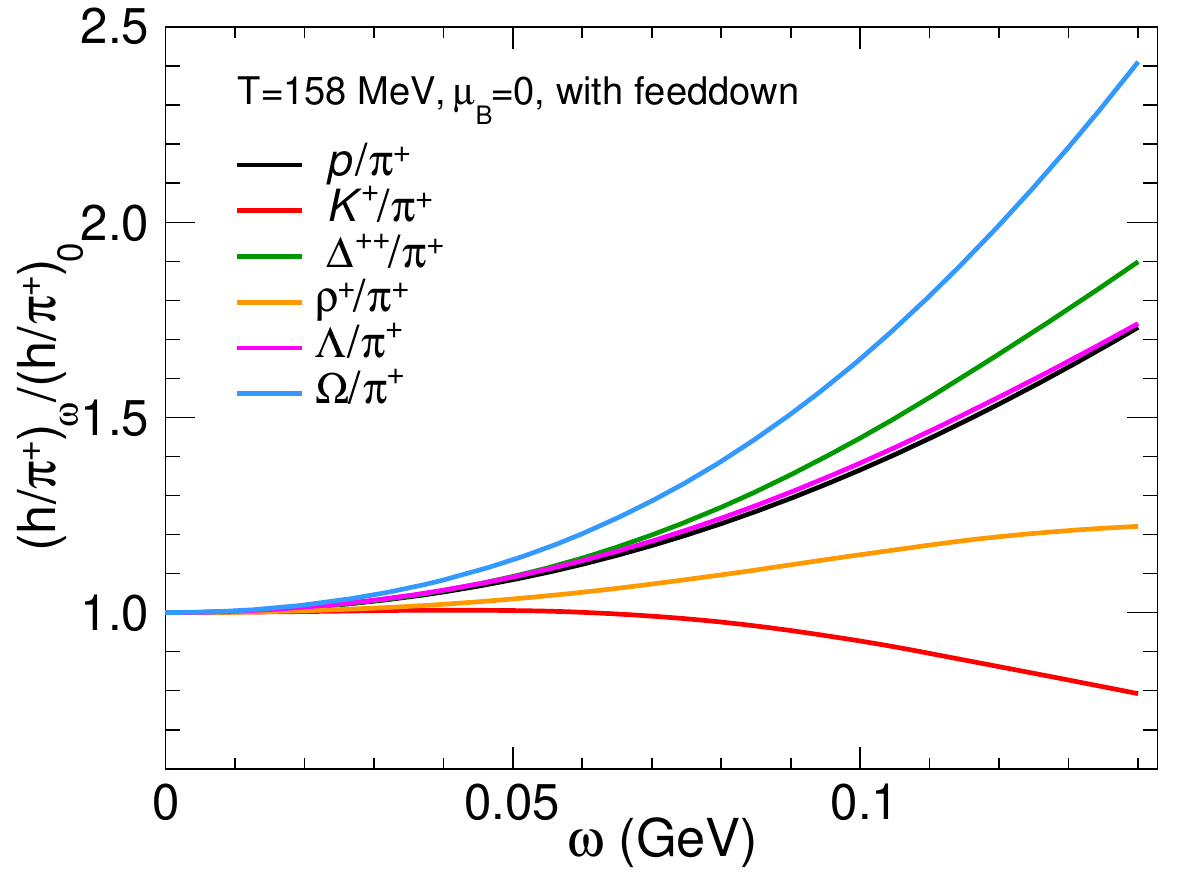}
    \caption{Dependence of various hadron-to-pion yield ratios (normalized by 
their corresponding values without rotation) on rotation, $\omega$, in HRG model 
at temperature $T=158$ MeV and vanishing chemical potential.}
    \label{fig4}
\end{figure}

Figure \ref{fig2} shows the relative variation in the number densities of 
different hadrons as a function of the rotational angular velocity ($\omega$). 
The results are normalized to their corresponding values in the absence of 
rotation. These calculations are performed at a temperature $T=158$ MeV and 
baryon chemical potential $\mu_B=0$, corresponding to the chemical freeze-out 
conditions in Pb$+$Pb collisions at the LHC~\cite{Andronic:2017pug}. In the 
figure, the dashed curves represent the relative changes in the primordial 
number densities, whereas the solid curves account for the total yields, 
including feed-down contributions arising from strong and weak decays.

The primordial $\pi^+$ yield decreases with increasing rotation, whereas the 
$K^+$ yield exhibits relatively weak sensitivity to rotation. However, after 
incorporating feed-down contributions, the number densities of both particles 
increase with rotation. In the case of $\Delta(1232)^{++}$, a strong
enhancement in the number density is observed as the rotational strength
increases. This enhancement arises from the reduction in the effective
mass and chemical potential associated with its large spin ($s=3/2$) and
electric charge ($+2$). Further, in addition to the higher spin, the number
density of higher mass particles such as $\Delta^{++}$ and $\Omega$ in the denominator
have relatively smaller number densities in the absence of rotation
compared to lighter hadrons. Consequently, the normalized density ratio
shows a larger relative increase for $\Delta^{++}$ and $\Omega$, indicating that rotational
effects preferentially enhances the production of higher
mass and higher spin particles.

Similarly, the primordial proton number density is also strongly affected by 
rotation, and the effect becomes more pronounced after including feed-down 
contributions, indicating substantial contributions from the decay of 
$\Delta(1232)^{++}$. The number density of $\rho^+$ increases significantly with 
rotation after including the contributions from decays. These results demonstrate
that rotation can induce 
significant modifications in hadron number densities. Furthermore, a clear 
mass ordering is observed, with heavier hadrons showing a stronger sensitivity 
to rotation. Hence, if significant vorticity persists until the freeze-out 
stage in heavy-ion collisions, it can have a substantial impact on the final 
hadron yields.

Figure~\ref{fig3} illustrates the dependence of the number densities of
various particles on rotation, normalized to their densities without
rotation at two different temperatures, 158 MeV and 100 MeV. In case of
$\pi^+$ and $K^+$, the normalized number densities at $T$ = 158 MeV are higher than
at $T$ = 100 MeV, indicating that the number densities increases with both
temperature and rotation. Conversely, for $p$, $\Delta^{++}$, $\rho^+$, and $\Omega$,
the trend is reversed with the ratios at $T$ = 100 MeV exceeding those
at $T$ = 158 MeV, despite the increase in individual number densities of 
these hadrons with temperature. The relative increase in the number densities 
of $p$, $\Delta^{++}$, $\rho^+$, and $\Omega$ with rotation is higher at temperature $100$ MeV.

Figure \ref{fig4} presents the relative variation of the number densities 
of different hadron-to-pion yield ratios due to rotation, normalized to their 
corresponding values in the absence of rotation, as a function of $\omega$. 
Most hadron-to-pion ratios exhibit a strong dependence on $\omega$. The
$K^+/\pi^+$ ratio shows decreasing trend as a function of rotation as compared to
without presence of rotation. A 
clear mass ordering is observed in the behavior of these ratios with increasing 
rotation, where heavier hadrons show a more pronounced enhancement. In 
particular, $\Omega/\pi^+$ and $\Delta^{++}/\pi^+$ ratio demonstrate strong sensitivity 
to rotation among the considered hadrons, owing to the large mass and higher 
spin.

\begin{figure}[htbp]
    \centering
    \includegraphics[width=\linewidth]{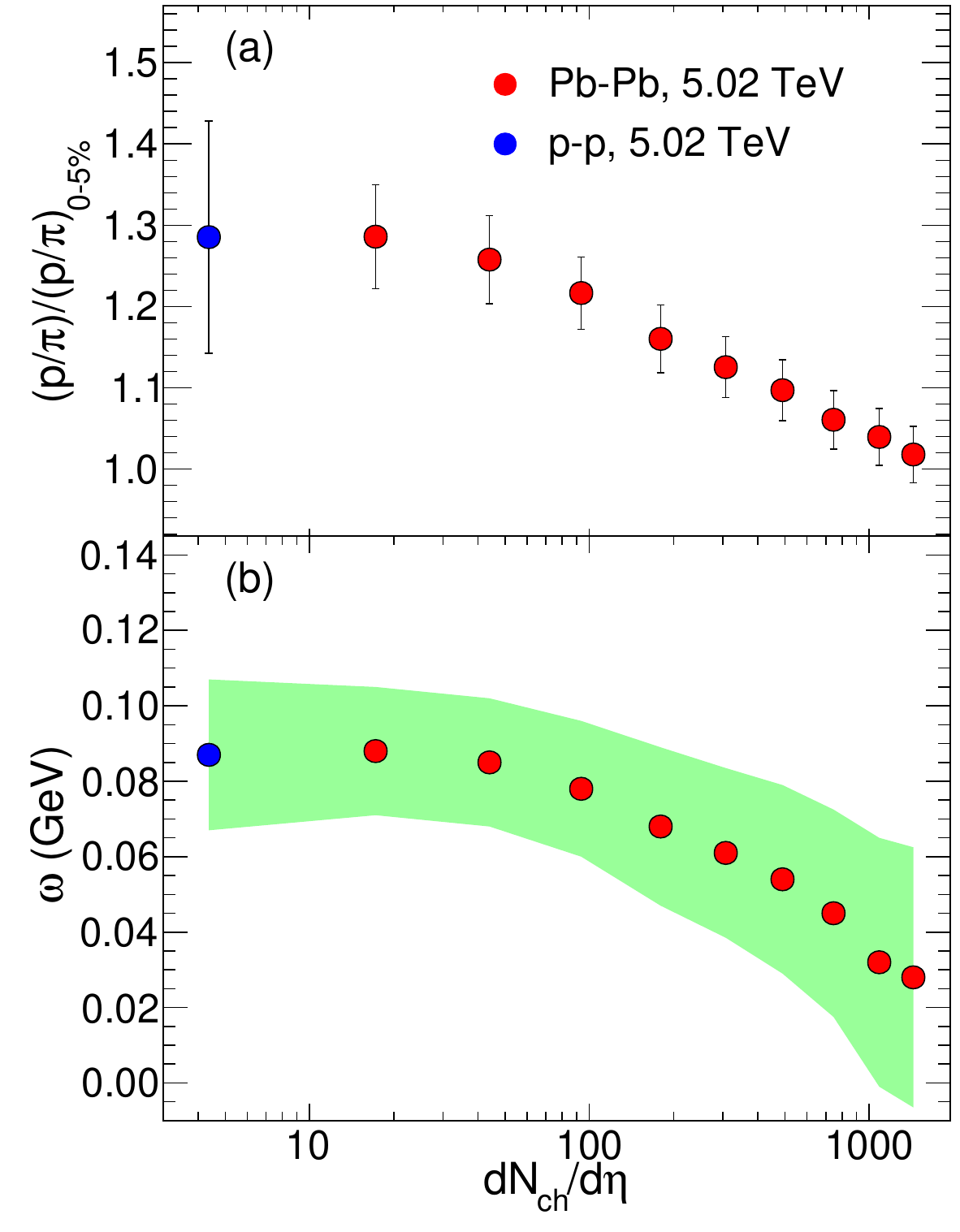}
    \caption{Top: Multiplicity dependence of the $p/\pi$ ratio (normalized by 
the $p/\pi$ ratio in $0-5\%$ collisions) measured by the ALICE Collaboration in 
Pb+Pb collisions at $\sqrt{s_{NN}}=5.02$ TeV. Bottom: Multiplicity dependence of 
rotation in HRG model extracted from experimental data to match the enhancement 
of the $p/\pi$ ratio relative to central collisions. The uncertainties associated 
with the $p/\pi$ ratio are propagated to estimate the $\omega$ are shown in shaded band.}
    \label{fig5}
\end{figure}
\subsection{Extraction of vorticity ($\omega$) from the $p/\pi$ ratios}
The deconfinement region of the QCD phase diagram for hot and dense matter
under the influence of global rotation as observed in heavy-ion collisions,
exhibits several intriguing features. In earlier study, it has been shown
that the combined effects of magnetic field and finite angular velocity 
can significantly enhance the reduction of the critical temperature, $T_c$ 
induced by the baryon chemical
potential~\cite{Padhan:2026mwg,Mukherjee:2023qvq}. In this context, 
it is important to investigate the effects of the magnetic field ($eB$) and 
angular velocity ($\omega$) on freeze-out parameters, in addition to baryon 
chemical potential ($\mu_B$) and temperature ($T$). Such a study is expected
to provide deeper insights into the thermodynamic properties and evolution of
the strongly interacting matter in presence of rotation, produced in heavy-ion
collisions.
One possible approach to probe the effects of global rotation is through
the centrality dependence of hadron yield ratio, specifically
proton-to-pion ($p/\pi$) ratio, which can serve as a sensitive observable
to study the role of rotation at chemical freeze-out.
In the present work, we estimate the magnitude of $\omega$ at chemical 
freeze-out in Pb$+$Pb collisions at LHC energies. A substantial value of 
$\omega$ at freeze-out could leave a noticeable impact on other experimental 
observables, for instance inducing a difference in the polarization of 
$\Lambda$ and $\bar{\Lambda}$ hyperons~\cite{STAR:2017ckg}. To quantify the rotational strength,
we correlate the relative enhancement 
of the $p/\pi$ ratio at different collision centralities with finite $\omega$, 
we provide an estimate of the rotational strength associated with each 
centrality class of the heavy-ion collisions from available experimental data.
Figure \ref{fig5} (a) shows the measurements by the ALICE collaboration of the 
$(p+\bar p)/(\pi^++\pi^-)$ ratio in Pb$+$Pb collisions at $\sqrt{s_{NN}}=5.02$ TeV, normalized 
to the corresponding value for the most central collisions. The experimental
data exhibit a systematic decrease in $p/\pi$ ratio from peripheral to central
collisions~\cite{ALICE:2019hno}. The uncorrelated uncertainties are also
depicted along with the data points. The suppression of the $p/\pi$ ratio
with increasing centrality is attributed to baryon annihilation processes
during the hadronic 
phase. In contrast, the vorticity $\omega$ is expected to be smaller in central 
collisions compared to peripheral collisions. Although, there are different mechanisms
responsible for the centrality dependence of $p/\pi$ ratio, we assume the
presence of vorticity as one possible contributing factors. As shown in
Fig~\ref{fig4}, a nonzero
vorticity leads to an enhancement of the $p/\pi$ ratio. We extract the corresponding 
values of vorticity ($\omega$) from the measured centrality dependence of the $p/\pi$ ratio.
For each centrality class, the $p/\pi$ ratios, normalized to the
corresponding values of the $0-5\%$ centrality interval, is mapped with
the one given by nonzero vorticity values shown in Fig~\ref{fig4}.
This correspondence enables the extraction of the vorticity strength
at a given centrality. This idea is based on the assumption that 
$0-5\%$ centrality corresponds to a system with negligible vorticity.
\begin{figure}[htbp]
    \centering
    \includegraphics[width=\linewidth]{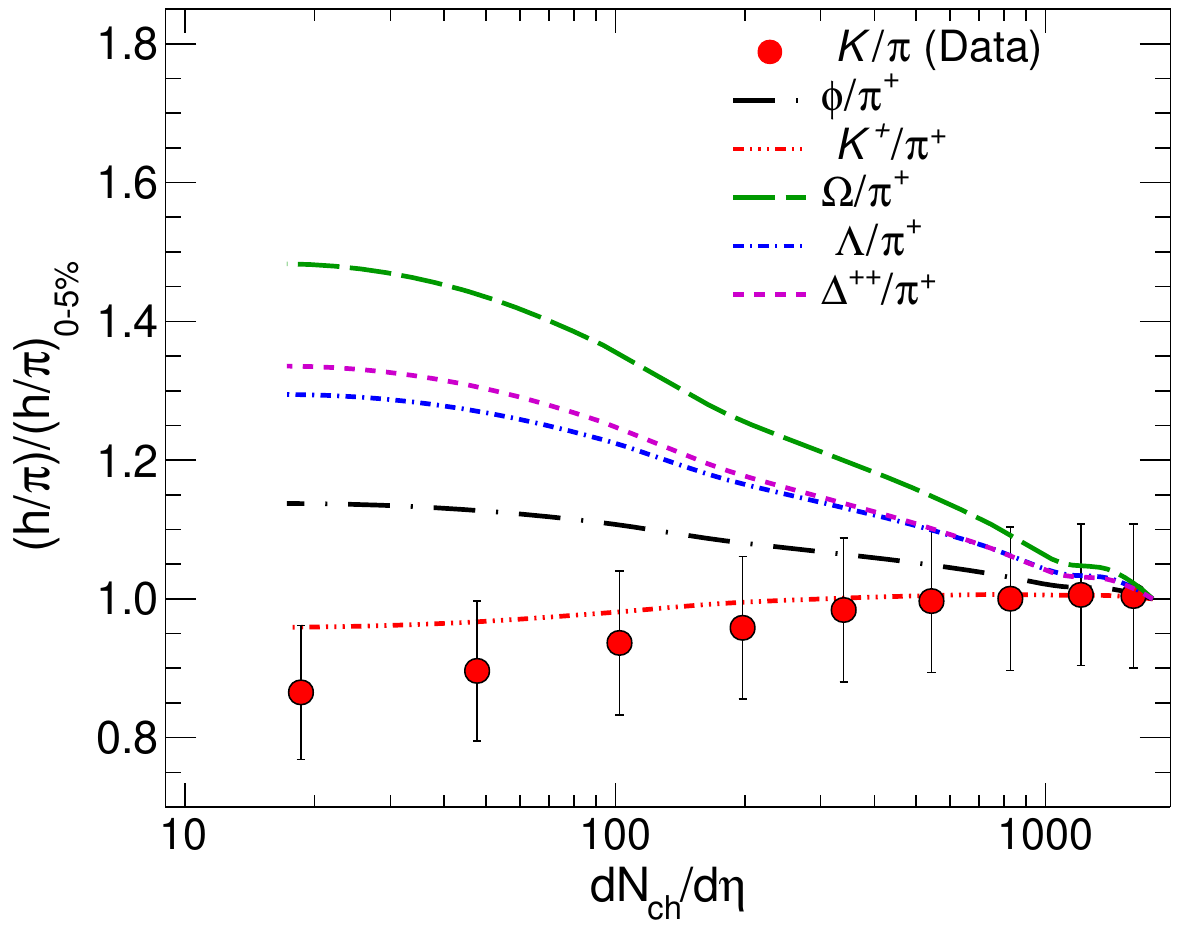}
    \caption{Multiplicity dependence of hadron yields 
      $\Phi/\pi^+$, $K^+/\pi^+$, $\Omega/\pi^+$, $\Lambda/\pi^+$, 
      and $\Delta^{++}/\pi^+$ in Pb+Pb collisions at LHC energies calculated in the 
HRG model at $T=158$ MeV and $\mu_B=0$. The ratios are normalized by their 
values in central collisions. } 
    \label{fig6}
\end{figure}

Figure \ref{fig5}(b) depicts the centrality dependence of rotation 
($\omega$) extracted from experimental data. The rotational strength is found 
to be maximal in peripheral collisions and decreases systematically towards 
central collisions. The upper limit of rotation, in peripheral Pb+Pb collisions, 
is estimated to be $0.088$ GeV or $\sim 10^{23}$ s$^{-1}$. For the most central case, 
the upper limit of rotation is obtained to be $0.028$ GeV or $\sim 10^{22}$ s$^{-1}$. 

It is important to note certain limitations associated with 
the present analysis. First, the chemical freeze-out temperature has 
been fixed at $T_{ch}=158$ MeV, although it may itself be modified in the 
presence of rotation. Second, the extracted $\omega$ is assumed to 
survive until the chemical freeze-out stage, thereby neglecting any
possible evolution or dissipation of the system's initial vorticity
during its dynamical expansion. These assumptions introduce systematic 
uncertainties in the extracted values of ($\omega$). Despite these caveats, the 
present study provides a reasonable estimate of the rotational magnitude at 
freeze-out, suggesting that rotation may significantly influence the hadron 
yields and potentially explain the observed centrality dependence of the
$p/\pi$ yield ratio.
\subsection{Effect of rotation on other hadron yield ratios}
Beyond the $p/\pi$ ratios, the centrality dependence of various hadron yield 
ratios can serve as an additional probe to asses the potential impact of 
rotation at freeze-out. Figure~\ref{fig6} illustrates the multiplicity 
dependence of $\Phi/\pi^+$, $K^+/\pi^+$, $\Omega/\pi^+$, $\Lambda/\pi^+$, 
      and $\Delta^{++}/\pi^+$
along with the experimental data for $(K^++K^-)/(\pi^++\pi^-)$ in Pb $+$ Pb collisions at 
$\sqrt{s_{NN}}=5.02$ TeV for comparison~\cite{ALICE:2019hno}.
The ratios are normalized to their corresponding values in the absence of 
rotation for the most central collisions.
The rotation has weak dependence on the kaon yields, hence the
normalized ratio is close to unity in the HRG model calculation.
The $K^+/\pi^+$ ratio has weak dependence on rotation which explains the 
experimental data. 
The ratios decrease from peripheral to central collisions owing to the dependency 
on rotation. The effect of rotation on higher mass particles is prominent in the 
centrality corresponding to peripheral collision.
A distinct spin dependence is evident, 
particularly for $\Omega^-$ hyperon in comparison with the other 
hyperons.

\begin{figure}[htbp]
    \centering
        \includegraphics[width=\linewidth]{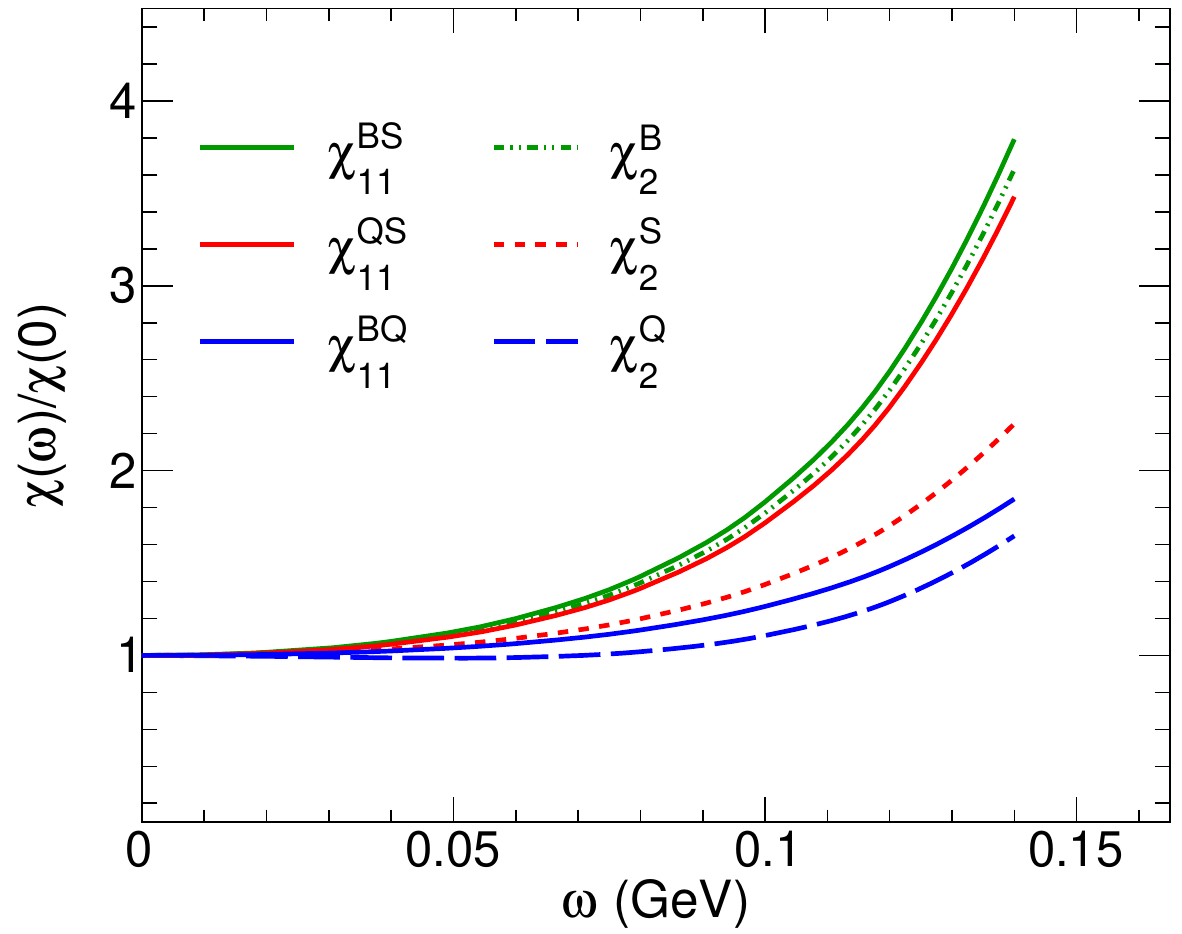}
    \caption{The dependence of quadratic susceptibilities of conserved charges 
    ($B$, $Q$, and $S$) and their correlations on rotation at temperature $T=158$ MeV and $\mu_B=0$. 
The susceptibilities are normalized by their respective values without rotation.}
    \label{fig7}
\end{figure}

\begin{figure}[htbp]
    \centering
        \includegraphics[width=\linewidth]{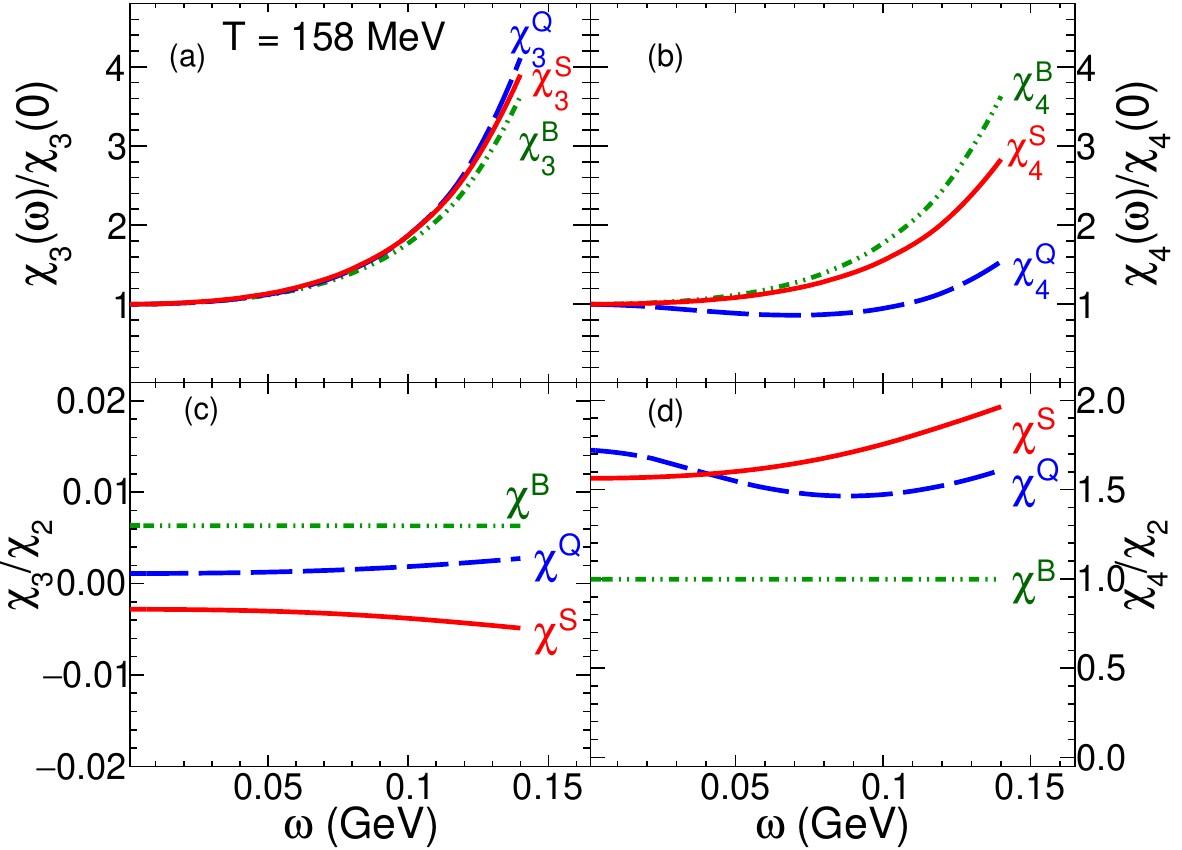}
    \caption{(top):The dependence of normalized higher order susceptibilities ($\chi_3$ in (a) and 
    $\chi_4$ in (b)) of the conserved charges on rotation. 
    (bottom): The dependence of third-to-second order susceptibility ratio, $\chi_3/\chi_2$,
    and fourth-to-second order susceptibility ratio, $\chi_4/\chi_2$, on rotation evaluated at $T=158$ MeV and $\mu_B=0$.} 
    \label{fig8}
\end{figure}

\subsection{Effect of rotation on fluctuations}
In HRG model calculations, the susceptibilities are observables sensitive to
the quantum numbers of the medium constituents. Thus, they can be used to
identify the contributions of different species of particles to QCD
thermodynamics~\cite{Ejiri:2006}.
In the present study we investigate the effect 
of rotation on conserved charge susceptibilities and the associated measurable 
fluctuations of hadrons. These observables serve as potential probes to 
quantify the magnitude of rotation in heavy-ion collisions by facilitating a 
direct comparison between theoretical predictions and experimental data.
The generalized susceptibilities of individual particle are defined as~\cite{Ding:2025nyh}:
\begin{equation}\label{e5}
    \chi_n^j=\frac{\partial^n (p/T^4)}{\partial(\mu_j/T)^n}
\end{equation}
where, $n$ represents the order of derivative, and $j$ denotes the 
corresponding conserved quantum numbers, namely baryon number ($B$), 
electric-charge ($Q$), and strangeness ($S$). 

The correlations among different conserved number serve as probes of the
QCD structure at finite temperature. 
These correlations of conserved charges in a thermalized medium are
quantified by the generalized susceptibilities as:
\begin{equation}\label{e5}
    \chi^{ij}_{11}=\frac{\partial^2 (p/T^4)}{\partial(\mu_i/T)\partial(\mu_j/T)}
\end{equation}
where $i,j \in$ \{$B,Q,S$\}. 
The quantities $\chi_{11}^{BQ}$, $\chi_{11}^{BS}$, and
$\chi_{11}^{QS}$ are sensitive to changes in baryon and strangeness
degrees of freedom~\cite{Koch:2005vg,Jeon:2000wg,Majumder:2006nq,HotQCD:2012fhj}.

Figure~\ref{fig7} shows the rotation dependence of the quadratic fluctuations and
their correlations of conserved quantities. The susceptibilities are normalized 
to their corresponding values without taking rotation ($\omega$ = 0) into account. 
The second-order susceptibilities
($\chi_2^B$, $\chi_2^Q$, and $\chi_2^S$) show the contributions from
individual conserved quantities, whereas the mixed susceptibilities take
the correlation of different conserved quantities into account.
The rapid increase of second-order susceptibilities is driven by the increase of hadron
number densities in HRG model, which are strongly dependent on $\omega$.
The $\chi_2^B$ and $\chi_{11}^{BS}$ show the maximum sensitivity, which is attributed
to the contributions from higher spin and heavier baryons.

Figure~\ref{fig8}, (a) and (b) show the rotation dependence of
higher order susceptibilities ($\chi_3$ and $\chi_4$) normalized with their
corresponding values without rotation. 
The third order susceptibilities ($\chi_{3}$) of the baryon, charge, and 
strangeness show similar dependency on rotation, where as the fourth-order 
susceptibility ($\chi_4$) of baryon number show more sensitivity to rotation as 
compared to strangeness and electric-charge.
Fig.~\ref{fig8} (c) and (d) show the susceptibility ratios 
($\chi_3/\chi_2$ and $\chi_4/\chi_2$) as a function of $\omega$. The ratios
$\chi_3/\chi_2$ for the conserved numbers remain nearly insensitive to rotation, despite
the strong  rotational dependence observed in the individual susceptibilities.
The net-strangeness susceptibility ratio exhibit a small dependence on
$\omega$, which becomes more discernible at higher values of $\omega$.
The ratio $\chi_4/\chi_2$ for net-baryon remains constant, whereas
net-electric charge and strangeness show dependence on $\omega$.

\begin{figure}[h]
    \centering
        \includegraphics[width=\linewidth]{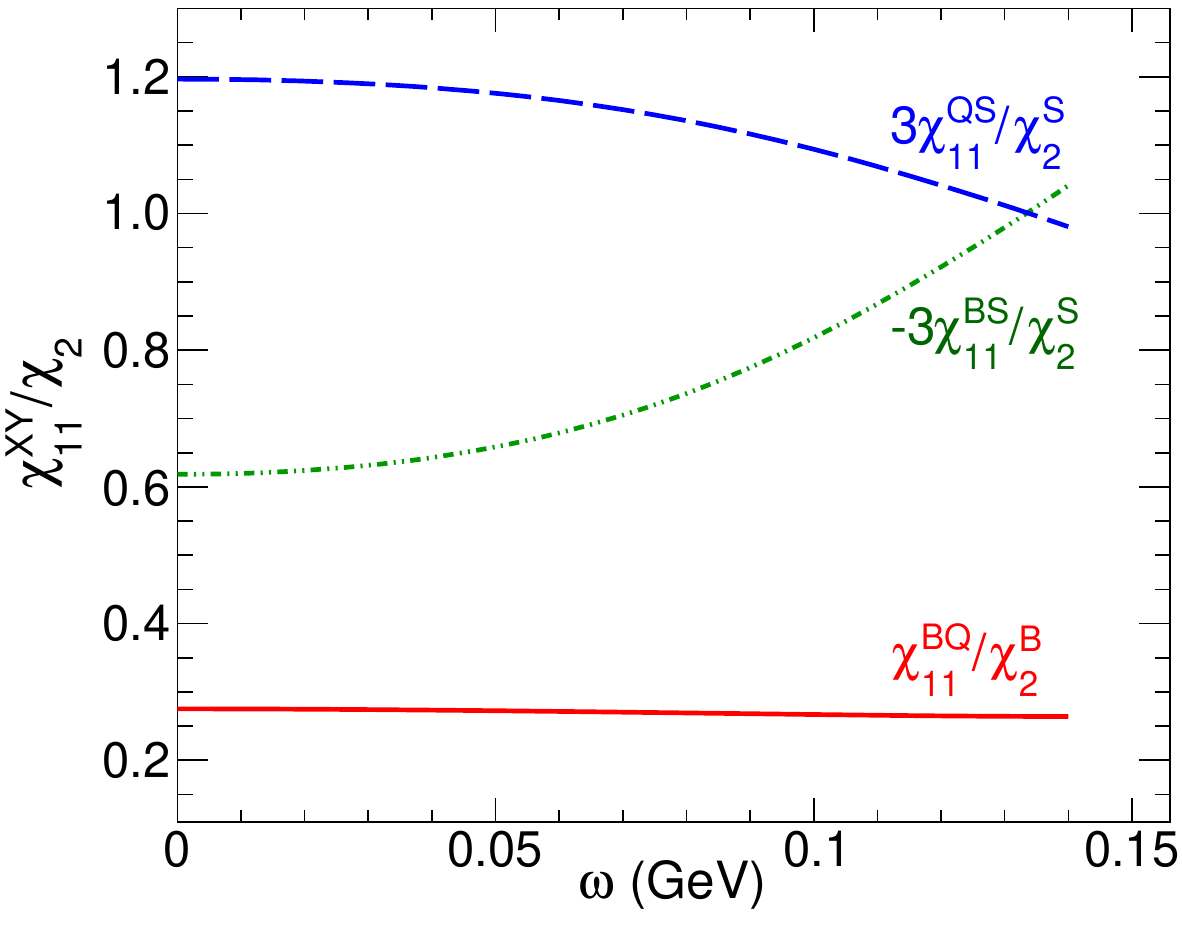}
    \caption{The dependence of $\chi_{11}^{XY}/\chi_2$ 
    on rotation evaluated at $T=158$ MeV and $\mu_B=0$.} 
    \label{fig9}
\end{figure}

\begin{figure}[h]
    \centering
        \includegraphics[width=\linewidth]{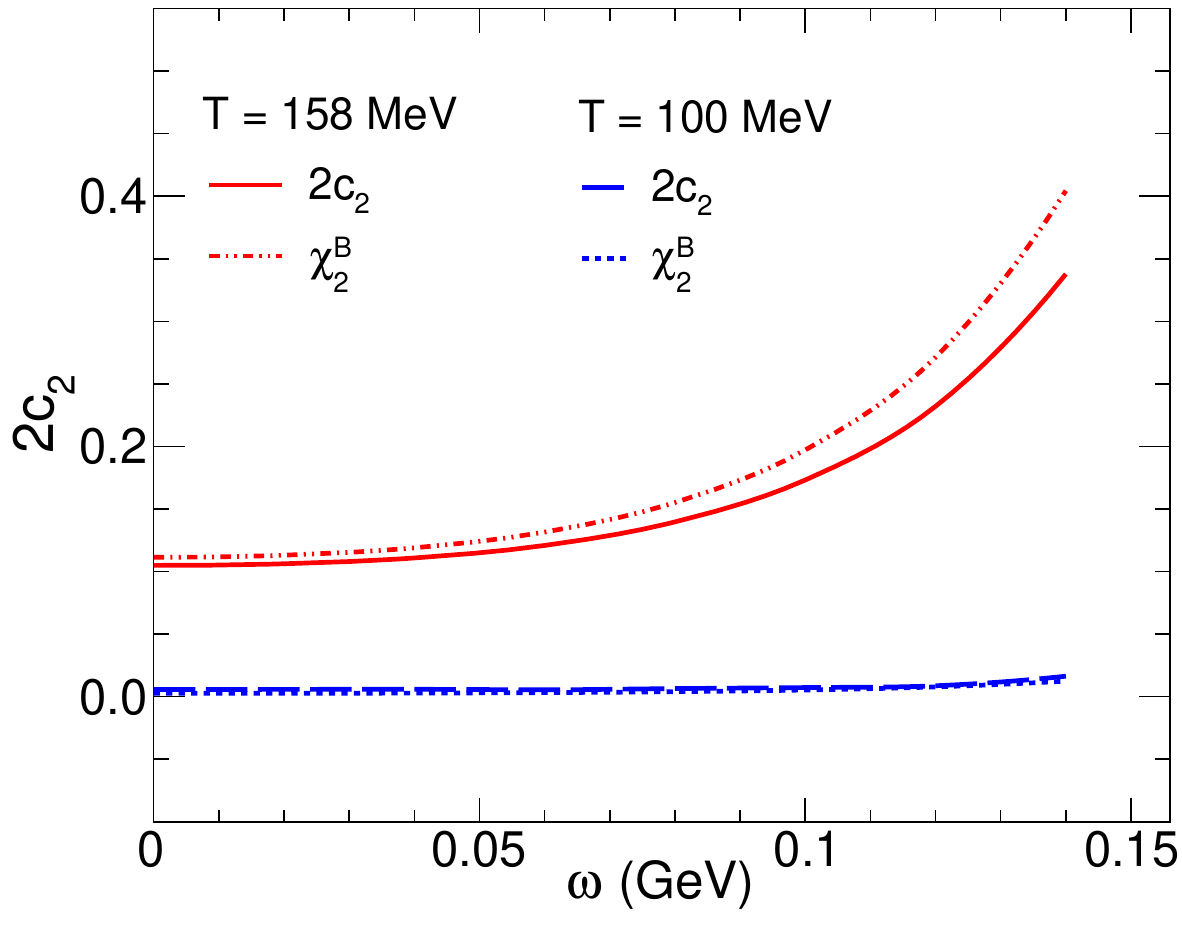}
    \caption{The dependence of $2c_2$ and $\chi_2^B$ 
    on rotation evaluated at $T=158$ MeV and $100$ MeV and $\mu_B=0$.} 
    \label{fig10}
\end{figure}

In order to understand whether the strangeness degrees-of-freedom in the
QGP can be described by weakly interacting quasi-quarks, we study
correlations of net strangeness fluctuations with fluctuations of net baryon
number and electric charge ($\chi_{11}^{BS}/\chi_2^S$ and $\chi_{11}^{QS}/\chi_2^S$).
The leading order perturbative correction can be eliminated by forming suitable
combination of ratios, such that those observables can be used to understand
experimental data on conserved charge fluctuations. This quantity is important
as leading order corrections do not cancel completely due to
differences in masses of light and strange quarks~\cite{HotQCD:2012fhj}. The 
observables -3$\chi_{11}^{BS}/\chi_2^S$,
3$\chi_{11}^{QS}/\chi_2^S$, and $\chi_{11}^{BQ}/\chi_2^B$ provide an 
useful diagnostic tool for identifying strong correlations between quarks and anti-quarks
~\cite{Koch:2005vg,HotQCD:2012fhj}.
Figure~\ref{fig9} shows the ratios of correlated and individual second-order
susceptibilities as a function of $\omega$. 
$3\chi_{11}^{QS}/\chi_2^S$ decreases with rotation, while $-3\chi_{11}^{BS}/\chi_2^S$ increases, 
symmetrically with the former since $\chi_{11}^{QS}/T^2 \sim -\chi_{11}^{BS}/T^2$ at high temperatures~\cite{HotQCD:2012fhj}.
The effect of rotation is
negligible on $\chi_{11}^{BQ}/\chi_2^B$ ratio. This is because the different masses of the 
light and strange quarks prevent the leading perturbative corrections from canceling.

At very small baryon density, the pressure can be expanded as a series
in $\mu_B/T$:
\begin{equation}
\frac{p}{T^4} \approx c_0 + c_2\left(\frac{\mu_B}{T}\right)^2 + c_4\left(\frac{\mu_B}{T}\right)^4 + c_6\left(\frac{\mu_B}{T}\right)^6 + O\left(\left(\frac{\mu_B}{T}\right)^8\right).
\end{equation}
The coefficient $c_0$ determines the pressure at $\mu_B$ = 0. The coefficients $c_2$,
$c_4$ and $c_6$ are related to the fluctuations $\chi_{ijk}^{BQS}$ of the conserved
charges $B$, $Q$ and $S$ at zero baryon density. The second coefficient $c_2$, which
is a mixture of second-order susceptibilities defined as~\cite{Marczenko:2024kko,Astrakhantsev:2024mat}:

\begin{equation}\label{e6}
   c_2 = \frac{1}{2}\chi_2^B+\frac{1}{3}\chi_{11}^{BS} + \frac{1}{18}\chi_2^S.
\end{equation}
The coefficient $c_2$ is calculated at $\mu_B$ = 0. For vanishing net-baryon
density, the strangeness neutrality condition can be approximated by
$\mu_S$ = 0, which leads to $\chi_{BB} \approx$ 2$c_2$. Figure~\ref{fig10}
shows the $c_2$ and $\chi_2^B$ dependence on rotation for two different
temperatures $T$ = 100 MeV and 158 MeV. It is observed that $\chi_2^B$ values
are systematically higher than 2$c_2$ for all values of $\omega$ at $T$ = 158 MeV.
There is a strong dependence of $\omega$ for different $T$.
For higher temperature $c_2$ and $\chi_2^B$ increases rapidly as a
function of $\omega$, whereas for $T$ = 100 MeV both remain constant
as a function of $\omega$. 

\section{Summary}\label{summary}

In this study, we investigated the impact of rotation on hadron yields
and their ratios within a modified HRG model. Our results demonstrate
that rotation can significantly modify hadron number densities,
due to the singular behavior in their dispersion relations.
The effect is particularly pronounced for higher-spin and higher-mass hadrons,
such as $\Omega$, $\Delta^{++}$, proton, and $\rho^+$ meson, while spin-zero
particles, including $\pi^+$ and $K^+$, exhibit comparatively
weaker sensitivity. Furthermore, feed-down contributions from resonance
decays enhance the rotational effects on the final hadron abundances.
An attempt is made to extract the upper limit of rotation ($\omega$)
from the ALICE experimental data. Assuming the enhancement of $p/\pi$
ratio is attributed to rotation, an upper limit of $\omega\sim$
$0.028$ GeV ($\sim10^{22}$ s$^{-1}$) for central collisions and $0.088$ GeV 
($\sim 10^{23}$ s$^{-1}$) for peripheral Pb+Pb
collisions at $\sqrt{s_{NN}}$  = 5.02 TeV is obtained.
The influence of rotation becomes more pronounced in peripheral
heavy-ion collisions, where larger vorticities are expected, and
gradually diminishes with increasing centrality.
These findings suggest that vorticity can play a significant role in
shaping the hadronic composition of the system at freeze-out and
should therefore be incorporated into thermal-model analyses of
particle production.
We have also studied the effect of rotation on susceptibilities of
different conserved quantities. There is a strong enhancement of
individual as well as correlated susceptibilities with rotation.
The susceptibility ratios $\chi_3/\chi_2$ for different conserved numbers are 
independent of rotation, whereas $\chi_4/\chi_2$ show rotation dependence except net-baryon.
The observables $3\chi_{11}^{QS}/\chi_2^S$ and $-3\chi_{11}^{BS}/\chi_2^S$ show 
equal and opposite dependence on $\omega$, while $\chi_{11}^{BQ}/\chi_2^B$ remains 
independent of $\omega$. The quadratic fluctuation of conserved baryon number increases 
with rotation at high temperature.

The inclusion of rotation can be considered as an additional
thermodynamic parameter, which may improve the extraction of freeze-out
conditions across different collision energies and centralities.
Our results provide valuable insights into the rotational
properties of strongly interacting matter and contribute to
future exploration of the QCD phase structure, in conjunction with
ongoing and forthcoming heavy-ion collision experiments. This study
provides a novel approach to probing the vortical structure of the 
medium created in heavy-ion collisions.

\nocite{*}
\bibliographystyle{apsrev4-1}
\bibliography{rotw}     

\end{document}